%% file: main.tex
\documentclass[conference]{IEEEtran}
\IEEEoverridecommandlockouts
\usepackage{cite}
\usepackage{amsmath,amssymb,amsfonts}
\usepackage{algorithmic}
\usepackage{graphicx}
\usepackage{textcomp}
\usepackage{xcolor}
\usepackage{hyperref}
\usepackage{lipsum}
\usepackage{tcolorbox}

\def\BibTeX{{\rm B\kern-.05em{\sc i\kern-.025em b}\kern-.08em
    T\kern-.1667em\lower.7ex\hbox{E}\kern-.125emX}}

\begin{document}

\title{An Empirical Study on Workflows and Security Policies in Popular GitHub Repositories}

\author{\IEEEauthorblockN{Jessy Ayala and Joshua Garcia}
\IEEEauthorblockA{\textit{Donald Bren School of Information and Computer Sciences}\\
\textit{University of California, Irvine, USA}\\
\{jessya1, joshug4\}@uci.edu}
}

\maketitle

\begin{abstract}
In open-source projects, anyone can contribute, so it is important to have an active continuous integration and continuous delivery (CI/CD) pipeline in addition to a protocol for reporting security concerns, especially in projects that are widely used and belong to the software supply chain. Many of these projects are hosted on GitHub, where maintainers can create automated workflows using GitHub Actions, introduced in 2019, for inspecting proposed changes to source code and defining a security policy for reporting vulnerabilities. We conduct an empirical study to measure the usage of GitHub workflows and security policies in thousands of popular repositories based on the number of stars. After querying the top one-hundred and top one-thousand repositories from all 181 trending GitHub topics, and the top 4,900 overall repositories, totaling just over 173 thousand projects, we find that 37\% of projects have workflows enabled and 7\% have a security policy in place. Using the top 60 repositories from each of the 34 most popular programming languages on GitHub, 2,040 projects total, we find that 57\% of projects have workflows enabled and 17\% have a security policy in place. Furthermore, from those top repositories that have support for GitHub CodeQL static analysis, which performs bug and vulnerability checks, only 13.5\% have it enabled; in fact, we find that only 1.7\% of the top repositories using Kotlin have an active CodeQL scanning workflow. These results highlight that open-source project maintainers should prioritize configuring workflows, enabling automated static analysis whenever possible, and defining a security policy to prevent vulnerabilities from being introduced or remaining in source code. \\
\end{abstract} 

\begin{IEEEkeywords}
empirical, GitHub, open-source, security, software supply chain
\end{IEEEkeywords}

\input{content/introduction.tex}
\input{content/related_work.tex}
\input{content/methodology.tex}
\input{content/results_and_discussion.tex}
\input{content/limitations_and_futurework.tex}
\input{content/conclusion.tex}

\input{content/references.tex}
\end{document}

%% file: content/introduction.tex
\section{Introduction}

With the explosion of open-source software, software supply chain attacks are on the rise \cite{b1}. In fact, many open-source projects are not only used by the general audience, but some companies rely on open-source technologies to build products \cite{b2}. Hence, software vulnerability management is crucial to ensuring the safety of popular projects. Many open-source projects are hosted on GitHub, where anyone can contribute.

In software engineering, workflows are used as a method of automating code operations. GitHub workflows allow project maintainers to automatically perform a wide range of checks (e.g., unit tests, link verification) on proposed changes for various reasons, including security-oriented ones. Recently, GitHub workflows can be easily configured with CodeQL, a static analysis tool built by GitHub, for code scanning \cite{b3}, which is especially important for widely used projects since CodeQL also scans repository source code for publicly documented vulnerabilities and will report this information to project maintainers \cite{b4}. Enabling CodeQL code scanning creates a workflow to perform bug and security checks when a contributor wants to add new changes to the repository, thereby preventing ``bad code'' from being introduced.

GitHub also supports the adoption of security policies, which usually describe a process for reporting any potential vulnerabilities found \cite{b5}. Project maintainers can set this up by including a \textit{SECURITY.md} file at the project root. Having such a protocol in place is critical to the safety of popular projects to ensure patching happens on an ongoing basis. 

In this paper, we conduct a large-scale empirical study on GitHub workflows and security policies from the top repositories from trending topics and those with the most widely used programming languages, totaling 2,040 projects and 175,156 public repositories. We provide preliminary observations to address the following research questions:

\begin{description}
 \item[\textbf{RQ1}] What \% of the top trending and overall GitHub projects have workflows and/or a security policy?
  \item[\textbf{RQ2}] What \% of the top GitHub projects written in popular programming languages have workflows and/or a security policy?
  \item[\textbf{RQ3}] What \% of the top GitHub repositories that primarily use a CodeQL-supported language have CodeQL-enabled (i.e., to reveal bugs and vulnerabilities)?
\end{description}

%% file: content/related_work.tex
\section{Related Work}

Related work on GitHub workflows and their usage in open-source projects are primarily from vulnerability discovery and software development perspectives. Benedetti et al. define a security assessment methodology to identify weaknesses (e.g., too many permissions given) in GitHub workflows \cite{b6}. On the other hand, Koishybayev et al. developed a tool to identify exploitable GitHub workflows \cite{b7}. Furthermore, a previous empirical GitHub workflow study focusing on popular projects found that nearly 22\% of them used workflows, with 2.8 configured on average per project, and investigated their impact on software development actions, such as commit frequency and issue resolution \cite{b8}. 

Now that GitHub workflows have been around for just over three years, our dataset consists of current popular repositories, instead of a general subset between a specified time period \cite{b9}, as of December 2022. We measure the usage of and investigate workflows from a vulnerability management lens by also checking whether CodeQL static analysis is used, when applicable, since project maintainers will be notified about bug and security findings. We also measure the usage of and collect GitHub security policies, where there has not been any previous work that we are aware of. 

%% file: content/methodology.tex

\section{Methodology}

We use the \textit{gh} command line tool \cite{b10}, developed by GitHub, to query for the top 1,000 repositories, based on the number of stars, from each of the 181 trending topics \cite{b11}, and the top 4,900 repositories overall \cite{b12}, totaling 173,116 repositories after removing projects that were either made private or taken down by GitHub (e.g., due to copyright issues). We also collect the top 60 repositories per popular programming language on GitHub, totaling 2,040 repositories, listed from the top GitHub Ranking list \cite{b13}. Note that though this source claims the top 100 repositories from each respective category, only 60 are available per language. 

From each repository, we forward their name and respective author, workflows, and security policies into a CSV file. We retrieve the contents of workflows and security policies with \textit{cURL} \cite{b14} and parse raw data with \textit{html2text} \cite{b15}. The data collected provides a mapping between the project owner and name to its workflows and security policy for reproducibility. We find that workflow metadata are structured by their title, whether they are active or not, and their corresponding id number from using the \textit{gh workflow list} command. We apply regular expressions using \textit{sed} to the contents of collected workflows and security policies to retain the CSV format.

We then use common data-science libraries implemented in Python (e.g., Pandas) to perform data analysis and create visualizations using the information we collected from each repository. All scripts, data scraped, and analyses are available on GitHub\footnote{https://github.com/jayala-29/svm2023-artifacts} to enable study and methodology reusability and replicability, and reproducibility.

%% file: content/results_and_discussion.tex
\section{Results and Discussion}

\subsection{RQ1: Projects from Trending Topics and Overall}

\begin{figure}[htp]
\centering
\begin{tabular}{@{}c@{}}
    \includegraphics[clip,width=0.87\columnwidth]{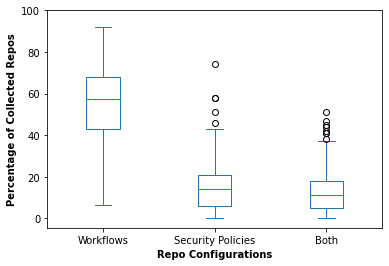} \\
    \small (a) Using the top 100 repos per trending topic (17,827 projects)
  \end{tabular}
  \setlength{\abovecaptionskip}{0pt}
  \begin{tabular}{@{}c@{}}
    \includegraphics[clip,width=0.87\columnwidth]{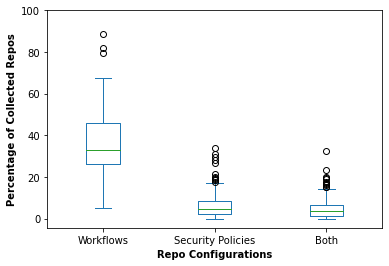} \\[\abovecaptionskip]
    \small (b) Using the top 1000 repos per trending topic (168,222 projects)
  \end{tabular}
  \vspace{\floatsep}
  \begin{tabular}{@{}c@{}}
    \includegraphics[clip,width=0.87\columnwidth]{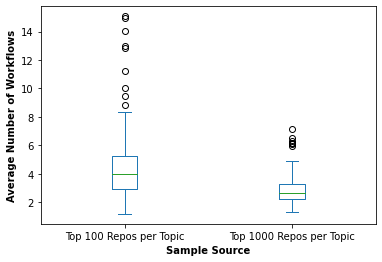} \\[\abovecaptionskip]
    \small (c) Average number of workflows per sample source
  \end{tabular}
\caption{Statistics from GitHub trending topics' repos with the most stars}
\end{figure}
 
In Fig. 1(a)-(b), we show box-and-whisker plots to summarize the percentages of repositories from 181 trending topics that use a workflow, have a security policy defined, and both. Each data point represents the percentage of repositories with a workflow enabled in their respective trending topic. In Fig. 1(a), the mean percentages are 55\%, 16\%, and 13\%, respectively. In Fig. 1(b), the mean percentages are 36\%, 7\%, and 5\%, respectively. We notice a decrease in all metrics with more repositories, which could be explained by including less popular projects, as pointed out by Chen et al \cite{b8}.

In Fig. 1(c), we show box-and-whisker plots to show statistics regarding the average number of workflows from trending topics for both sample sources. Each data point represents the average number of workflows in their respective trending topic. The overall means are 4.44 and 2.87 workflows per repository, with at least one workflow enabled. Like what is observed in Fig. 1(a)-(b), there is a noticeable decrease in the average number of workflows as the repositories sample size increases.

Based on the results, repositories with more stars have more workflows and security policies than those with fewer stars. This indicates that repositories with higher popularity would be more likely to have some software management process in place by means of an automated workflow to run checks on proposed changes or a security policy to receive bug reports in the case where someone finds a potential vulnerability. However, the lack of security policies in repositories is concerning because if there is no security policy defined, then users who find a vulnerability do not have a protocol to follow for reporting it to project maintainers for triaging purposes. 

We also notice a low percentage in popular repositories that have at least one workflow and a security policy defined. In the best case, projects have both. To address this issue, \textbf{project maintainers of popular repositories should consider adding a workflow or security policy, if either is missing, to prevent malicious actors from successfully inserting ``bad code'' via pull requests or exploiting a known vulnerability} (e.g., ones that could be detected through static analysis). In fact, some projects belonging to trending topics are critical to the software supply chain (e.g., npm code packages).

Though the results from popular repositories imply a lack in security policy usage, we discover progress made since Chen et al \cite{b8} discovered that nearly 22\% of popular projects used workflows, with 2.8 workflows configured on average per project as of February 2021. When only looking at workflows, we notice a 14\% increase in the number of repositories that use at least one workflow and a 0.07 increase in the average amount of workflows configured per project after querying just over 168,000 repositories from trending topics. 

\vspace{-0.25cm}

\begin{table}[htbp]
\caption{Statistics after Scraping Information from the Top 4,900 Repositories}
\centering
\begin{tabular}{ |p{0.21\textwidth}|p{0.04\textwidth}| } 
 \hline
 \textbf{Information Calculated} & \textbf{Value}  \\ 
  \hline
 \% of repos with workflows & 51\% \\ 
  \hline
 \% of repos with a security policy & 19\% \\ 
 \hline
 \% of repos with both & 16\% \\ 
 \hline
 Average number of workflows & 4.46 \\ 
 \hline
\end{tabular}
\end{table}
	
In Table 1, we present statistics about the top 4,900 repositories, consisting of 4,894 projects since 6 were either removed by GitHub or unavailable. These results are closely aligned with the information calculated in Fig. 1(a), which represents just under 17 thousand projects, as opposed to Fig. 1(b), which represents just over 168,000 projects. This indicates a similarity in the way project maintainers use or do not use workflows and security policies for their repositories.

Combining results from querying the top 1,000 projects from each trending topic and the top 4,900 projects, 37\% of projects have at least one workflow enabled, and 7\% of projects have a security policy defined. 
Given the low rate of defining security policies (7\%) compared to enabling a workflow (37\%), \textbf{there is an opportunity for researchers to support project maintainers by providing tool support for defining security policies for their projects or studying what is preventing wider adoption of security policy creation.}


\subsection{RQ2: Projects Written in Popular Programming Languages}

\begin{figure}[htp]
\centering
\begin{tabular}{@{}c@{}}
    \includegraphics[clip,width=0.87\columnwidth]{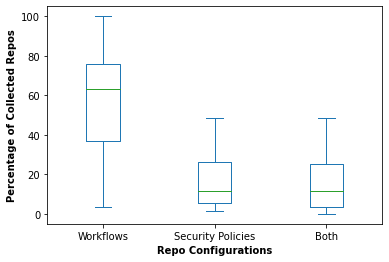} \\
    \small (a) Using the top 34 programming languages (2040 projects)
  \end{tabular}
  \begin{tabular}{@{}c@{}}
    \includegraphics[clip,width=0.87\columnwidth]{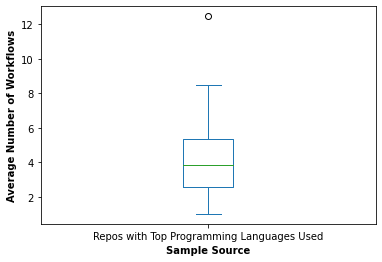} \\
    \small (b) Average number of workflows 
    
  \end{tabular}
\caption{Statistics from GitHub repos written in the top 34 programming languages with the most stars} 
\vspace{-4ex}
\end{figure}
 
In Fig. 2(a), we show a box-and-whisker plot to summarize the percentage of repositories from the top 34 programming languages that use a workflow, have a security policy defined, and both. The mean values are 57\%, 17\%, and 15\%, which are closely aligned with the data in Table 1, showing information calculated using the top 60 repositories per programming language, respectively, totaling 2,040 projects. In Fig. 2(b), we show another box-and-whisker plot to show statistics describing the average number of workflows from the same sample source. The overall mean is 4.24 workflows per repository in projects with at least one workflow enabled, and all categories' average mean is greater than zero.

Based on the results presented, they are again much closer to the statistics shown in Fig. 1(a), indicating  a similarity in the way project maintainers use or do not use workflows and security policies for their repositories. \textbf{Researchers should further investigate a larger sample of repositories for these languages to study how these values change---and whether or not at least one of the workflows is security-related when a security policy is not used.} These practices are essential for popular repositories containing software instead of projects with only resources or documents (e.g., lecture materials).

Because this is our only sample source where all projects queried are software-based, we want to also point out the progress made in adopting GitHub workflows since their inception in 2019. We again would like to compare against the results from Chen et al \cite{b8}. Almost two years later, as of December 2022, we see a 35\% increase in workflow usage and a 1.44 increase in the average amount of workflows configured per project written in the most widely used programming languages on GitHub. \textbf{There is progress in adopting GitHub workflows, which is likely to continually increase as projects become more popular and code size grows.}
 
\subsection{RQ3: Projects with CodeQL Support}

\vspace{-0.35cm}

\begin{table}[htbp]
\caption{Top GitHub Repositories Primarily Written in a CodeQL-Supported Language}
\begin{center}
    \begin{tabular}{ |p{0.13\textwidth}|p{0.19\textwidth}| } 
 \hline
 \textbf{CodeQL-supported Language} & \textbf{\% of Repos with a CodeQL Workflow Enabled}  \\ 
  \hline
 C & 10\% \\ 
  \hline
 C++ & 10\% \\ 
 \hline
 C\# & 10\% \\ 
 \hline
 Go & 35\% \\ 
 \hline
  Java & 11.7\% \\ 
  \hline
 JavaScript & 15\% \\ 
 \hline
 Kotlin & 1.7\% \\ 
 \hline
 Python & 10\% \\ 
 \hline
 Ruby & 21.7\% \\ 
 \hline
 TypeScript & 10\% \\ 
 \hline
\end{tabular}
\end{center}
\end{table}

\vspace{-0.35cm}

In Table 2, we calculate the percentage of the most popular CodeQL-supported repositories, based on the number of stars, with at least the CodeQL workflow enabled. This is especially important for vulnerability management since CodeQL scans code for general bugs and vulnerabilities. We show these metrics to highlight the underuse of enabling a CodeQL workflow in projects where it is supported.

For instance, of the top 60 Kotlin repositories, only 1.7\% have an active CodeQL workflow running. On the other hand, of the top 60 Go repositories, 35\% have an active CodeQL workflow running, the greatest percentage out of the category of languages explored. These numbers are especially concerning since enabling CodeQL in a CodeQL-supported repository is relatively simple and provides detailed alerts to project maintainers \cite{b4}. If an adversary were to run CodeQL on repository source code and vulnerabilities are detected, they would be informed of them before project maintainers would.

    Overall, we calculate that of the top 60 repositories per programming language with CodeQL support, only 13.5\% of them have an active CodeQL workflow running. \textbf{We advise open-source project maintainers to configure code scanning whenever possible for early vulnerability triaging, even more so for projects that are widely used.}

%% file: content/limitations_and_futurework.tex
\section{Limitations and Future Work}

In our data collection, we relied on our references for querying purposes \cite{b11}, \cite{b12}, \cite{b13}. Therefore, our sample sizes may seem odd (e.g., scraping information from the top 60 repositories primarily written in a popular programming language, 34 distinct). In addition, we did not remove duplicates if they showed up from different sample sources, and we did not check if repositories are forks of one another. 

A threat to external validity is our focus on a single static analysis tool, CodeQL. We focus on CodeQL due to its ease of integration with projects where it is supported, but understand that there are other security code scanning tools available. 

Future work includes continuing exploring other categories of popular repositories (i.e., collect more data), looking deeper into collected security policies (e.g., similarities in word usage), determining whether workflows are security related, and including a human-centered aspect to understand why security policies are not widely used in popular open-source projects.

%% file: content/conclusion.tex
\section{Conclusion}



In this paper, we present progress toward measuring the usage of software vulnerability management features in popular open-source projects available on GitHub. After querying the top one-hundred and top one-thousand repositories from all 181 trending GitHub topics, and the top 4,900 repositories, totaling just over 173,000 projects, we find that 37\% of projects have workflows enabled and 7\% have a security policy in place. Using the top 60 repositories from each of the 34 most popular programming languages on GitHub, 2,040 projects total, we find that 57\% of projects have workflows enabled and 17\% have a security policy in place. Furthermore, from those repositories that have support for CodeQL static analysis to perform bug and security checks, only 13.5\% have this feature enabled. Though results indicate an increase in the usage of workflows in popular GitHub repositories based on findings from prior work, security policies are still in progress of being widely adopted. 

%% file: main.bbl
\begin{thebibliography}{00}
\bibitem{b1} J. Hutchings, The importance of improving supply chain security in open source, \href{https://github.blog/2022-11-09-improving-open-source-supply-chain-security}{https://github.blog/2022-11-09-improving-open-source-supply-chain-security}, Accessed: 2022-11-20, November 2022.
\bibitem{b2} M. Woodward, Octoverse 2022: 10 years of tracking open source software, \href{https://github.blog/2022-11-17-octoverse-2022-10-years-of-tracking-open-source}{https://github.blog/2022-11-17-octoverse-2022-10-years-of-tracking-open-source}, Accessed: 2022-11-28, November 2022.
\bibitem{b3} GitHub, (2018) GitHub CodeQL (Version 2.11.2) [Source code]. \href{https://github.com/github/codeql}{https://github.com/github/codeql}.
\bibitem{b4} GitHub, Code scanning, \href{https://docs.github.com/en/code-security/code-scanning/automatically-scanning-your-code-for-vulnerabilities-and-errors/about-code-scanning}{https://docs.github.com/en/code-security/code-scanning/automatically-scanning-your-code-for-vulnerabilities-and-errors/about-code-scanning}, Accessed: 2022-11-24, November 2022.
\bibitem{b5} GitHub, Adding a security policy to your repository, \href{https://docs.github.com/en/code-security/getting-started/adding-a-security-policy-to-your-repository}{https://docs.github.com/en/code-security/getting-started/adding-a-security-policy-to-your-repository}, Accessed: 2022-11-20, November 2022.
\bibitem{b6} G. Benedetti, L. Verderame, and A. Merlo, ``Automatic Security Assessment of GitHub Actions Workflows,'' In \textit{Proceedings of the 2022 ACM Workshop on Software Supply Chain Offensive Research and Ecosystem Defenses (SCORED'22)}, 2022, pp. 37–45, doi: \href{https://doi.org/10.1145/3560835.3564554}{https://doi.org/10.1145/3560835.3564554}.
\bibitem{b7} I. Koishybayev et al, ``Characterizing the Security of Github CI Workflows,'' In \textit{Proceedings of the 31st USENIX Security Symposium (USENIX Security'22)}, 2022, pp. 2747-2763.
\bibitem{b8} T. Chen, Y. Zhang, S. Chen, T. Wang, Y. Wu, ``Let’s Supercharge the Workflows: An Empirical Study of GitHub Actions,'' In \textit{Proceedings of the 2021 IEEE International Conference on Software Quality, Reliability, and Security Companion (QRS-C'21)}, 2021, pp. 1089-1098, doi: \href{https://doi.org/10.1109/QRS-C55045.2021.00163}{https://doi.org/10.1109/QRS-C55045.2021.00163}.
\bibitem{b9} A. Decan, T. Mens, P. R. Mazrae and M. Golzadeh, ``On the Use of GitHub Actions in Software Development Repositories,'' In \textit{Proceedings of the 2022 IEEE International Conference on Software Maintenance and Evolution (ICSME'22)}, Limassol, Cyprus, 2022, pp. 235-245, doi: \href{https://doi.org/10.1109/ICSME55016.2022.00029}{https://doi.org/10.1109/ICSME55016.2022.00029}.
\bibitem{b10} D. Stenberg, (2015) cURL (Version 7.82.0) [Source code]. \href{https://github.com/curl/curl}{https://github.com/curl/curl}.
\bibitem{b11} GitHub, Popular topics on GitHub, \href{https://github.com/topics}{https://github.com/topics}, Accessed 2022-12-10, December 2022.
\bibitem{b12} T. Kokubun, Gitstar Ranking, \href{https://gitstar-ranking.com/repositories}{https://gitstar-ranking.com/repositories}, Accessed: 2022-12-11, December 2022.
\bibitem{b13} E. Li, GitHub Ranking, \href{https://github.com/EvanLi/Github-Ranking}{https://github.com/EvanLi/Github-Ranking}, Accessed: 2022-12-12, December 2022.
\bibitem{b14} GitHub, (2020) GitHub CLI (Version 2.18.1) [Source code]. \href{https://github.com/cli/cli}{https://github.com/cli/cli}.
\bibitem{b15} A. Swartz, (2011) html2text (Version 2.1.1) [Source code]. \href{https://github.com/aaronsw/html2text}{https://github.com/aaronsw/html2text}.
\end{thebibliography}
